\documentclass[aps,cpl,reprint,preprintnumbers,superscriptaddress,amsmath,amssymb,bibnotes,longbibliography]{revtex4-1}

\usepackage{graphicx}
\usepackage{dcolumn}
\usepackage{bm}
\usepackage[pdfstartview=FitH, CJKbookmarks=true, bookmarksnumbered=true, bookmarksopen=true, colorlinks, pdfborder=001, linkcolor=blue, anchorcolor=blue, citecolor=blue, urlcolor=blue]{hyperref}

\begin{document}
	\title{Ce-site dilution in the ferromagnetic Kondo lattice CeRh$_6$Ge$_4$}
	\author{Jia-Cheng Xu}
	\affiliation{Center for Correlated Matter and Department of Physics, Zhejiang University, Hangzhou 310058, China}
	\author{Hang Su}
	\affiliation{Center for Correlated Matter and Department of Physics, Zhejiang University, Hangzhou 310058, China}
	\author{Rohit Kumar}
	\affiliation{Center for Correlated Matter and Department of Physics, Zhejiang University, Hangzhou 310058, China}
	\author{Shuai-Shuai Luo}
	\affiliation{Center for Correlated Matter and Department of Physics, Zhejiang University, Hangzhou 310058, China}
	\author{Zhi-Yong Nie}
	\affiliation{Center for Correlated Matter and Department of Physics, Zhejiang University, Hangzhou 310058, China}
	\author{An Wang}
	\affiliation{Center for Correlated Matter and Department of Physics, Zhejiang University, Hangzhou 310058, China}
	\author{Feng Du}
	\affiliation{Center for Correlated Matter and Department of Physics, Zhejiang University, Hangzhou 310058, China}
	\author{Rui Li}
	\affiliation{Center for Correlated Matter and Department of Physics, Zhejiang University, Hangzhou 310058, China}

	\author{Michael Smidman}
	\affiliation  {Center for Correlated Matter and Department of Physics, Zhejiang University, Hangzhou 310058, China}
	\affiliation  {Zhejiang Province Key Laboratory of Quantum Technology and Device, Department of Physics, Zhejiang University, Hangzhou 310058, China}
	
	\author{Hui-Qiu Yuan}
	\email[Corresponding author: ]{hqyuan@zju.edu.cn}
	\affiliation  {Center for Correlated Matter and Department of Physics, Zhejiang University, Hangzhou 310058, China}
	
	\affiliation  {Zhejiang Province Key Laboratory of Quantum Technology and Device, Department of Physics, Zhejiang University, Hangzhou 310058, China}
	\affiliation  {State Key Laboratory of Silicon Materials, Zhejiang University, Hangzhou 310058, China}

	\begin{abstract}
		The heavy fermion ferromagnet CeRh$_6$Ge$_4$ is the first example of a clean stoichiometric system where the ferromagnetic transition can be continuously suppressed by hydrostatic pressure to a quantum critical point. In order to reveal the outcome when the magnetic lattice of CeRh$_6$Ge$_4$ is diluted with non-magnetic atoms, this study reports comprehensive measurements of the physical properties of both single crystal and polycrystalline samples of La$_x$Ce$_{1-x}$Rh$_6$Ge$_4$. With increasing $x$, the Curie temperature decreases, and no transition is observed for $x$~$>$~0.25, while the system evolves from exhibiting coherent Kondo lattice behaviors at low $x$, to the Kondo impurity scenario at large $x$. Moreover, non-Fermi liquid behavior (NFL) is observed over a wide doping range, which agrees well with the disordered Kondo model for 0.52~$\leq$~$x$~$\leq$~0.66, while strange metal behavior is revealed in the vicinity of $x_c$~=~0.26.

	\begin{description}
		\item[PACS numbers]{71.27.+a, 75.40.-s, 72.15.Qm}
		
	\end{description}

	\end{abstract}
	
	\maketitle

	The highly tunable competition between the Ruderman--Kittel--Kasuya--Yosida (RKKY) interaction and the Kondo effect in heavy fermion systems makes them one of the most widely studied testbeds for quantum critical behavior in condensed matter physics \cite{1si2010heavy,2gegenwart2008quantum,38High-pressure,39Heavyfermionsmagneticfields,40TuningtheHeavyFermion,47Heavy}. Among the non-thermal tuning parameters, dilution of the magnetic Kondo lattice via substitution of non-magnetic atoms is an important method for tuning the magnetic transition \cite{36thompson2011holes,37kumar2014kondo}. In Ce-based compounds, La doping can be used to dilute the magnetic lattice composed of Ce ions. One interesting example is that of CePd$_{3}$ \cite{7lawrence1996kondo} where La doping leads to Kondo hole behavior. The physical properties of the heavy fermion compound CeNi$_{2}$Ge$_{2}$ evolves from coherent Fermi liquid (FL) behavior at low La doping, to a non-Fermi liquid (NFL) at intermediate doping levels and local Fermi liquid behavior at high La doping \cite{8pikul2012single,9pikul2010lack}. A similar evolution of the physical properties is observed in La-doped CeRhIn$_{5}$ \cite{10pagliuso2002site}.

	However, unlike antiferromagnetic quantum critical points (QCPs) \cite{1si2010heavy,2gegenwart2008quantum,38High-pressure,39Heavyfermionsmagneticfields,40TuningtheHeavyFermion,47Heavy}, where numerous cases have been found experimentally, ferromagnetic (FM) QCPs in clean systems have hardly been observed. Recently, it was found that CeRh$_6$Ge$_4$ belongs to a rare class of clean systems whose FM transition can be continuously suppressed by hydrostatic pressure to reveal an FM QCP \cite{13shen2020strange}. In CeRh$_6$Ge$_4$, strange metal behaviors such as a linear-$T$ resistivity, and a logarithmic divergence of the specific heat coefficient were observed close to the FM QCP \cite{13shen2020strange}. A subsequent study of quantum oscillations in CeRh$_{6}$Ge$_{4}$ demonstrated that the 4$f$ electrons remain localized at ambient pressure and hence do not contribute to the Fermi surface \cite{42wang2021localized}. An angle-resolved photoemission spectroscopy study at elevated temperatures provided spectroscopic evidence for the presence of anisotropic hybridization between the Ce-4$f$ and conduction elections in CeRh$_6$Ge$_4$ \cite{41wu2021anisotropic}, and it was inferred that the anisotropy of the low energy crystalline electric field (CEF) orbitals plays an important role in bringing about the anisotropic $c-f$ coupling \cite{43shu2021magnetic}. In such a heavy-fermion ferromagnetic material, Ce-site substitution by La will not only suppress the RKKY interaction but also weaken the Kondo effect, which motivates us to study the effects of La doping on the magnetism and the critical behavior of CeRh$_6$Ge$_4$.

	In this letter, the synthesis and physical properties of La$_x$Ce$_{1-x}$Rh$_6$Ge$_4$ are reported \cite{11vosswinkel2012bismuth,13shen2020strange}. Polycrystalline samples were synthesized by arc melting the constituent elements in a stoichiometric ratio. The as-cast samples were wrapped in Ta-foil, sealed in evacuated quartz ampoules and then annealed. Single crystals of La$_x$Ce$_{1-x}$Rh$_6$Ge$_4$ were grown using a Bi flux method \cite{11vosswinkel2012bismuth,13shen2020strange}. The elements were combined in a molar ratio of La$_x$Ce$_{1-x}$Rh$_6$Ge$_4$:Bi of 1:100, and sealed in an evacuated quartz tube. The tube was heated to 1400~K for 10-13~hours and then cooled to 773~K at 2~K/hour. The ampoules were then centrifuged to remove excess Bi. The remaining Bi on the surface of the crystals was removed by placing them in a 1:1 molar mixture of H$_2$O$_2$ and glacial acetic acid for several days. The crystal structure of the polycrystals was characterized by powder X-ray diffraction (XRD) using a Rigaku Ultima IV diffractometer with Cu-K$_{\alpha1}$ radiation. The composition of all doped samples was checked by energy-dispersive X-ray spectroscopy which was carried out using a Hitachi SU-8010 scanning electron microscope. Electrical transport and specific heat measurements were performed using a Quantum Design Physical Property Measurement System (PPMS).

		\begin{figure*}
		\includegraphics[angle=0,width=0.8\textwidth]{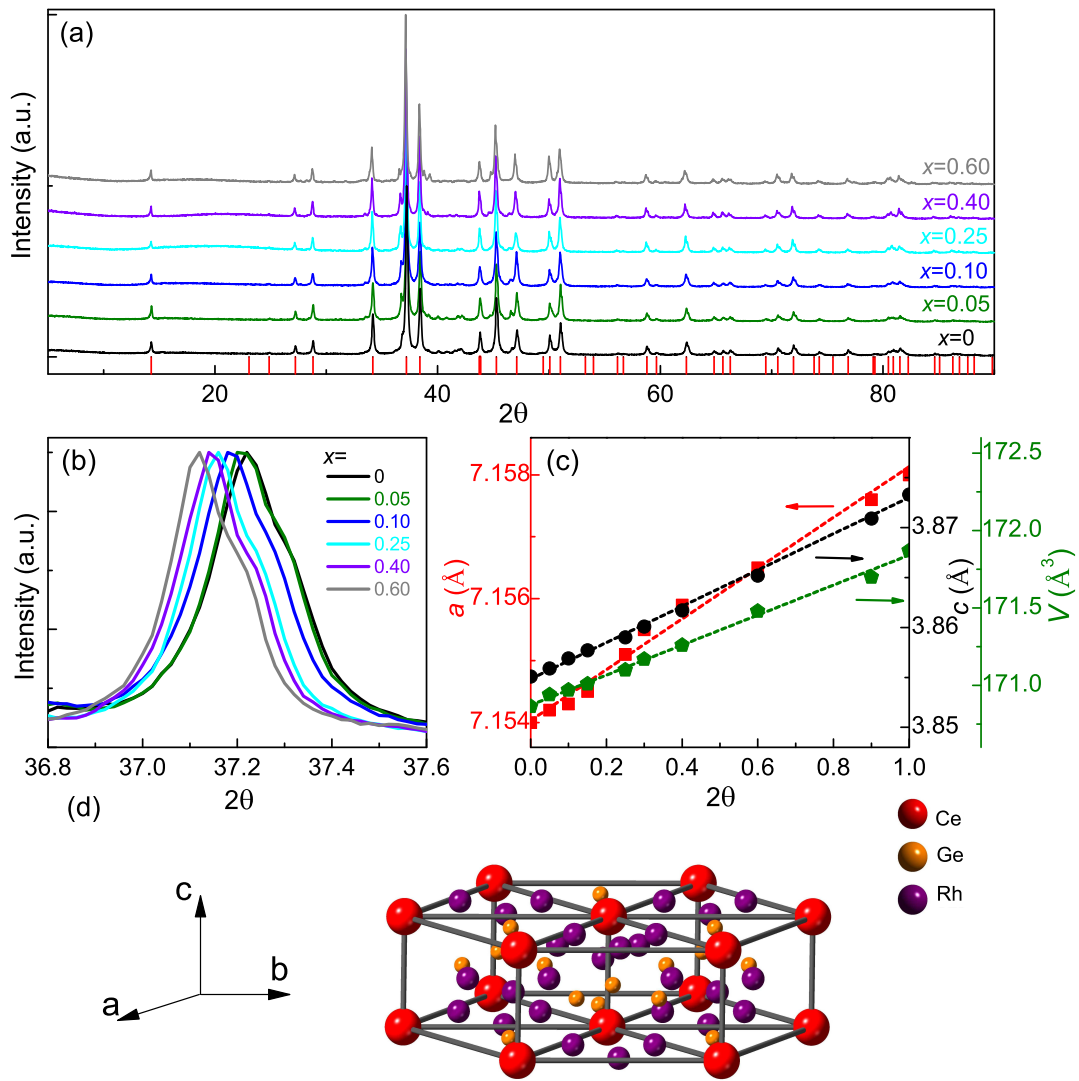}
		\vspace{-12pt} \caption{\label{figure1}(a) The XRD patterns of powdered polycrystals with $x$~=~0, 0.05, 0.10, 0.25, 0.40, 0.60. The peak positions corresponding to CeRh$_6$Ge$_4$ are marked by the red vertical lines. (b) Enlargement of the strongest Bragg peak, in which the peak position shifts to lower angle with increasing $x$. (c) The lattice constants $a$, $c$ and the unit cell volume $V$ of La$_x$Ce$_{1-x}$Rh$_6$Ge$_4$ versus $x$. (d) The crystal structure of the parent compound, CeRh$_6$Ge$_4$. }
		\vspace{-12pt}
	\end{figure*}

	XRD patterns of the powdered polycrystalline samples are shown in Fig.~\ref{figure1}(a). Here, the peak positions are compared to those expected for stoichiometric CeRh$_6$Ge$_4$, which has the LiCo$_{6}$P$_{4}$ type crystal structure \cite{14buschmann1991darstellung}. All doped compositions shown in Fig.~\ref{figure1}(a) have very similar Bragg positions to that of CeRh$_6$Ge$_4$ [P$\bar{6}$m2 (No. 187, D$^1_{3h}$)], suggesting that the doped samples also belong to the same crystal structure. After annealing, only a few very small peaks corresponding to impurity phases could be resolved. In Fig.~\ref{figure1}(b), the peaks with the highest intensity are displayed for all compositions, demonstrating that the peaks shift to lower angles as the La doping level is increased, suggesting the expansion of the lattice. Figure~\ref{figure1}(c) displays the variation of the lattice parameters $a$, $c$ and the volume of the unit cell $V$ as a function of La doping.

	\begin{figure}
		\includegraphics[angle=0,width=0.45\textwidth]{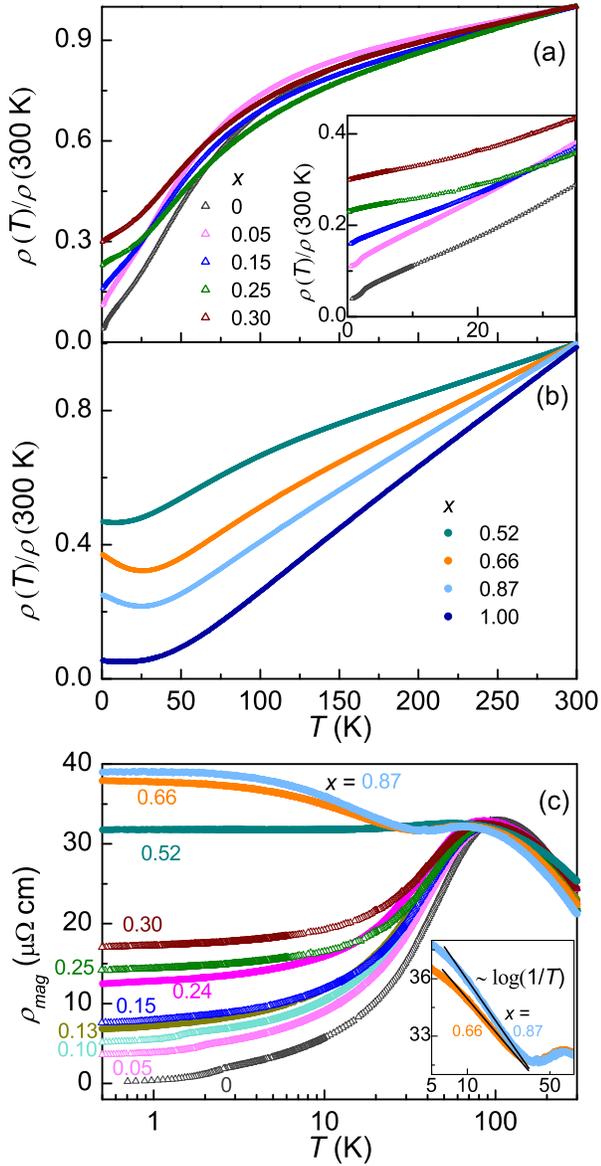}
		\vspace{-12pt} \caption{\label{figure2} Temperature dependence of the normalized resistivity for (a) polycrystalline Ce-rich compositions, and (b) La-rich single crystals. The inset is an enlargement for clarity. (c) Temperature variation of the magnetic contribution per Ce atom to the resistivity ($\rho_{mag}$) of all samples, where $\rho_{mag}$=($\rho_x-\rho_{La}$)/(1$-x$). Data for polycrystals and single crystals are represented by open triangles and filled circles, respectively. The inset is an enlargement showing the $\sim$log($1/T$) behavior of $\rho_{mag}(T)$ for $x$~=~0.66 and 0.87. }
		\vspace{-12pt}
	\end{figure}

	 The temperature dependence of the resistivity normalized by the value at 300~K is shown in Figs.~\ref{figure2}(a) and (b). Note that due to the poor quality of La-rich polycrystalline samples, for $x$~$>$~0.30 the data for single crystals is displayed. It is clear that the two sides of the series exhibit qualitatively different behaviors. Because of the disorder effect induced by elemental substitutions, the residual resistivity increases with doping on the Ce-rich side whereas, it gradually decreases for the La-rich side. The resistivity of La$_{0.66}$Ce$_{0.34}$Rh$_6$Ge$_4$ and La$_{0.87}$Ce$_{0.13}$Rh$_6$Ge$_4$ exhibits an upturn below 30~K. To study the doping evolution of the magnetic contribution to the resistivity per Ce ($\rho_{mag}$), the resistivity of LaRh$_6$Ge$_4$ was subtracted from the doped samples, which is defined as $\rho_{mag}$=($\rho_x-\rho_{La}$)/(1$-x$), and is displayed in Fig.~\ref{figure2}(c) for all compositions. In order to make sure that there is no fundamental difference between the physical properties of single and polycrystalline samples at the same composition, single crystals corresponding to low doping levels ($x$~=~0.13 and 0.24) were also grown and characterized. It is evident from the data of Ce-rich crystals ($x$~=~0.13 and 0.24), that the behavior is highly consistent between polycrystalline and single crystalline samples [see Fig.~\ref{figure2}(c)].

	All curves in Fig.~\ref{figure2}(c) exhibit a maximum near 100~K, which has also been observed in stoichiometric CeRh$_6$Ge$_4$ \cite{13shen2020strange}. Above the high temperature maximum ($T_{max}$), the $\rho_{mag}(T)$ of all samples behaves similarly, except for the value of the $T_{max}$, which shifts to slightly lower temperature with increasing $x$. The $T_{max}$ can be related to the combined effects of the low lying CEF excitations and the Kondo scattering. On the contrary, the behavior of $\rho_{mag}(T)$ below $T_{max}$ exhibits a strong composition dependence. In the Ce-rich region ($x$~$\leq$~0.30), $\rho_{mag}$ decreases significantly with decreasing temperature as a result of coherent Kondo lattice behavior. For $x$~=~0.52, after a slight decrease, the $\rho_{mag}$ becomes almost temperature independent down to 0.5~K while for $x$~=~0.66 and 0.87 there is a logarithmic divergence [$\rho_{mag}$~$\sim$~log(1/$T$)] from 20~K to 10~K, as shown in the inset. The low temperature log(1/$T$) behavior is attributed to incoherent Kondo scattering from the fully occupied CEF ground state, as is typical for Kondo impurity systems \cite{3bauer1991anomalous,22cox1988transport}, in contrast to the coherent scattering in the case of a well ordered Kondo lattice. In general, as $x$ increases, La$_x$Ce$_{1-x}$Rh$_6$Ge$_4$ evolves from Kondo lattice behaviors in the Ce-rich region to the Kondo impurity regime in the La-rich region. This also can be observed in other dilute Kondo systems, such as Ce$_x$La$_{1-x}$Cu$_6$ \cite{20onuki1985kondo}, Ce$_x$La$_{1-x}$Ni$_2$Ge$_2$ \cite{8pikul2012single,9pikul2010lack} and Ce$_x$La$_{1-x}$TiGe$_3$ \cite{44lee2019suppression}).

	\begin{figure}
		\includegraphics[angle=0,width=0.45\textwidth]{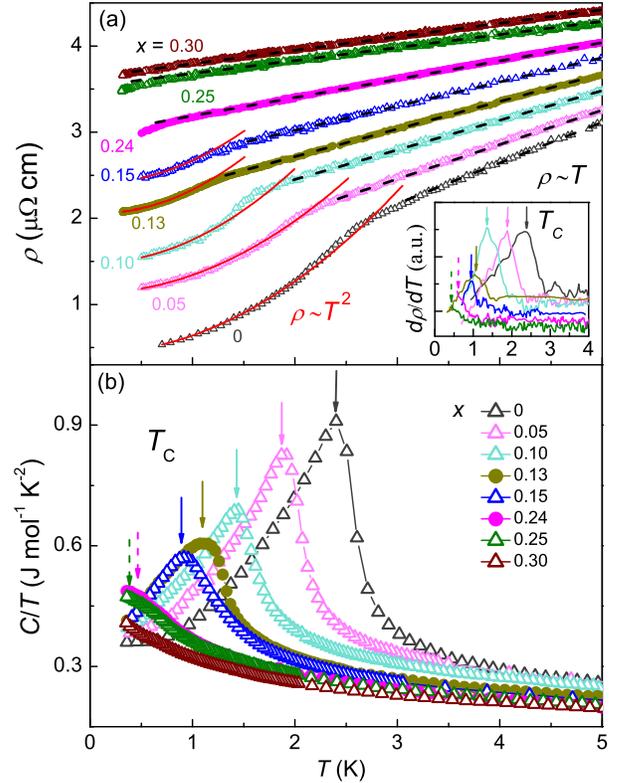}
		\vspace{-12pt} \caption{\label{figure3}Low temperature resistivity (a) and specific heat (b) for Ce-rich samples. Data for polycrystals and single crystals are represented by open triangles and filled circles, respectively. In (a), the resistivity curves are vertically shifted by 0.2~$\mu\Omega$~cm for clarity. The black dashed lines represent a linear fit to the data and the red solid curves show fits to a quadratic temperature dependence for a Fermi liquid, $\rho(T)$~$\sim$~$T^2$. The inset shows the derivative d$\rho$/d$T$ for samples with doping concentration $x$~$\leq$~0.25, where the peak position corresponds to $T_C$. In (b), solid arrows mark the positions of $T_C$, while the dashed arrows highlight the existence of a magnetic transition for the $x$~=~0.24 and 0.25 samples, while no transition is detected in $C/T$ for $x$~=~0.30. }
		\vspace{-12pt}
	\end{figure}

\begin{figure}
		\includegraphics[angle=0,width=0.45\textwidth]{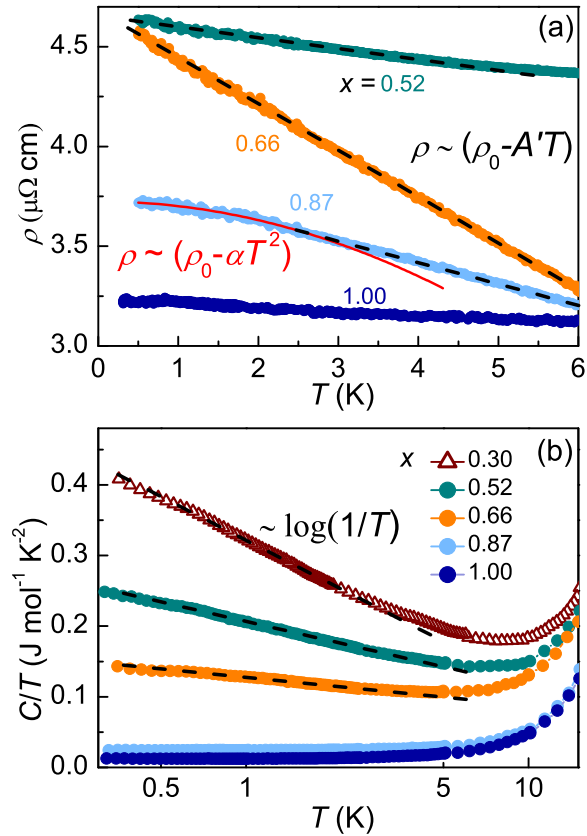}
		\vspace{-12pt} \caption{\label{figure4}(a) Low temperature resistivity for $x$~$\geq$~0.52. The data are all shifted vertically by 0.07 ~$\mu\Omega$~cm for clarity, except for $x$~=~0.52 which is shifted by 1.05~$\mu\Omega$~cm. The black dashed lines highlight a linear increase of $\rho(T)$ with the decreasing temperature. The red solid curve shows the fit of $\rho(T)$ to the local Fermi liquid behavior [$\rho$~$\sim$~($\rho_0-\alpha$$T^2$)] below 2.5~K for $x$~=~0.87. For $x$~=~1, the slight increase of $\rho(T)$ at low temperatures is likely attributed to a tiny amount of magnetic impurities. (b) Low temperature $C/T$ for $x$~$\geq$~0.30 on a logarithmic temperature scale. Data for polycrystals and single crystals are represented by open triangles and filled circles, respectively. The log($1/T$) behavior highlighted by the black dashed lines is observed in the doping range $x$~=~0.30~-~0.66. }
		\vspace{-12pt}
	\end{figure}

	The low temperature resistivity is shown in Fig.~\ref{figure3}(a), and the first derivative of the resistivity d$\rho$/d$T$ is displayed in the inset, which is utilized to define the Curie temperature $T_C$. A pronounced transition in d$\rho$/d$T$ [Fig.~\ref{figure3}(a)] and the specific heat [Fig.~\ref{figure3}(b)] is observed for compositions up to $x$~=~0.15, while for $x$~=~0.24 and 0.25 just an onset of magnetic transition is visible as a small hump in $C/T$ [Fig.~\ref{figure3}(a)] and as an upturn in d$\rho$/d$T$ [Fig.~\ref{figure3}(b)], but the determination of the Curie temperature $T_C$ for these compositions requires further measurements down to lower temperatures. In Fig.~\ref{figure3}, the behavior of $C/T$ at low doping levels is similar for polycrystals and single crystals ($x$~=~0.13 and 0.24), again showing that the properties of single crystals and polycrystalline samples are similar. Since the interaction between local magnetic moments is responsible for the ferromagnetic order in CeRh$_6$Ge$_4$, weakening of the RKKY interaction by La doping is the primary reason for the suppression of Curie temperature. A transition is not visible for the $x$~=~0.30 composition down to 0.3~K, and instead, the resistivity decreases linearly with decreasing temperature [Fig.~\ref{figure3}(a)], while $C/T$ increases [Fig.~\ref{figure4}(b)]. For $x$~$\leq$~0.3, the resistivity in the paramagnetic state shows non-Fermi liquid behavior which is manifested in the linear-$T$ dependent behavior of the resistivity (marked with black dashed lines), whereas in the magnetic state a quadratic dependence [$\rho(T)$=${\rho_0}$+$A$${T^2}$, marked with solid red lines] suggestive of coherent Fermi liquid behavior is observed.

	In Fig.~\ref{figure4}(a), low temperature transport and thermodynamic measurements of La-rich alloys are displayed. The black dashed lines are used to highlight the NFL behavior, where $\rho$~$\sim$~($\rho_0-A'T$) down to 0.3~K for $x$~=~0.52 and 0.66 but only down to 2.5~K for $x$~=~0.87. For the latter, the data below 2.5~K can be fitted with $\rho~\sim$~($\rho_0-\alpha$$T^2$), which is characteristic of incoherent (local) Kondo systems \cite{25nozieres1974fermi}. Therefore, similar to diluted Ce$_x$La$_{1-x}$Ni$_2$Ge$_2$ \cite{9pikul2010lack}, this diluted sample ($x$~=~0.87) exhibits local FL behavior below 2.5~K, suggesting a weak intersite coupling between the local 4$f$ moments.

	In Fig.~\ref{figure4}(b), the logarithmic temperature dependence (as shown by black dashed lines) of the specific heat [$C/T$~$\sim$~log($1/T$)] for $x$~=~0.3, 0.52 and 0.66 is displayed, while $C/T$ has no upturn and is almost temperature independent below 5~K for $x$~=~0.87. The behavior of the latter is similar to that observed in LaRh$_6$Ge$_4$, which is a paramagnetic metal without the Kondo effect. Hence, this temperature-independent behavior of $C/T$ implies the weak Kondo screening of local moments in the dilute limit \cite{26oliveira1981specific}. The logarithmic divergent $C/T$ along with the $T$-linear resistivity [$\rho$~$\sim$~$T$, as shown in Fig.~\ref{figure3}(a)] for $x$~=~0.30 is similar to the strange metal behavior seen under pressure in CeRh$_6$Ge$_4$ \cite{13shen2020strange}. For 0.66~$\geq$~$x$~$\geq$~0.52, the samples exhibit NFL behavior with $\rho$~$\sim$~($\rho_0-A'T$) and $C/T$~$\sim$~log($1/T$), which can be described within the framework of the disordered Kondo model \cite{32miranda1996kondo,28de1998evidence}.

	\begin{figure}
		\includegraphics[angle=0,width=0.45\textwidth]{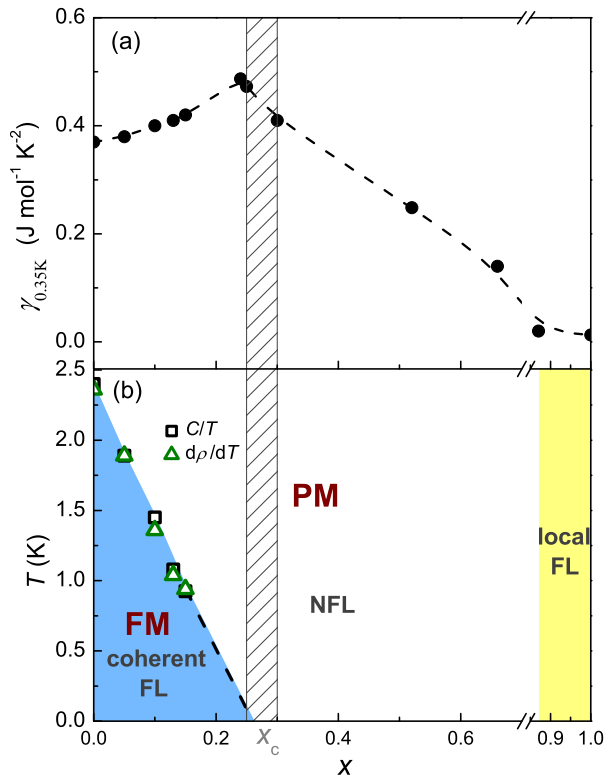}
		\vspace{-12pt} \caption{\label{figure5}(a) 
			Concentration dependence of the specific heat coefficient $\gamma_{0.35\text{K}}$ (as $C/T$ at 0.35~K). (b) $T$-$x$ phase diagram of La$_x$Ce$_{1-x}$Rh$_6$Ge$_4$. The open squares and triangles denote $T_C$ derived from the specific heat and d$\rho$/d$T$, respectively. The black dashed line is the extrapolation of the initial slope of $T_C$($x$), which reaches zero temperature at $x_c$~$\approx$~0.26. The shaded blue region corresponds to coherent Fermi liquid behavior and ferromagnetic order at low doping, while the local FL in the dilute limit is in yellow. NFL behavior is observed over a wide doping range (0.3~$\leq$~$x$~$\leq$~0.66). }
		\vspace{-12pt}
	\end{figure}

	In Fig.~\ref{figure5}(a), $\gamma_{0.35\text{K}}$ which is defined as $C/T$ at 0.35~K, is plotted against the doping concentration $x$. A pronounced maximum is observed at the composition $x$~$\approx$~0.24. It is noted that the values of $\gamma_{0.35\text{K}}$ for $x$~=~0.24 and 0.25 are taken from the points near the magnetic transition, which exceed the Sommerfeld coefficient at zero temperature. On the other hand, $\gamma_{0.35\text{K}}$ near $x_c$~=~0.26 is underestimated in comparison with the corresponding Sommerfeld coefficient. Therefore, the Sommerfeld coefficient is likely peaked near the critical concentration where the ferromagnetic order is suppressed. In Fig.~\ref{figure5}(b), the temperature-composition phase diagram derived from all the measurements is displayed. The transition temperatures determined from the specific heat and resistivity are in good agreement. For dense Kondo systems with low La-doping ($x$~$\leq$~0.15), coherent FL behavior is observed in the FM state, while for dilute Kondo alloys ($x$~$\geq$~0.87), the low-$T$ behavior can be described as a local Fermi liquid, represented by the yellow region. Similar to Ce$_x$La$_{1-x}$Ni$_2$Ge$_2$ \cite{8pikul2012single,9pikul2010lack}), these two FL regimes are separated by a NFL region (0.25~$<$~$x$~$\leq$~0.66), wherein the area near $x_c$~=~0.26 shows behaviors similar to the strange metal phase observed in the case of pressured induced FM QCP \cite{13shen2020strange}, while at higher doping the behavior corresponds to the disordered Kondo model.
	
	At low doping levels ($x$~$\leq$~0.15), the evolution of magnetism in La$_x$Ce$_{1-x}$Rh$_6$Ge$_4$ is similar to that of another dilute Kondo ferromagnet Ce$_x$La$_{1-x}$TiGe$_3$, where $T_C$ is linearly suppressed as $x$ increases \cite{44lee2019suppression}. If it is assumed that $T_C$ continues to decrease linearly with $x$, it will be completely suppressed near $x_c$~=~0.26, as highlighted by the black dashed line in Fig.~\ref{figure5}(b), which is close to where there is a maximum in the Sommerfeld coefficient. It is noted that a ferromagentic transition still survives for $x$~=~0.24 and 0.25 with an onset transition above 0.5~K. However, these two points are not shown in the phase diagram since their transition temperatures are below the temperature limit of measurements. It is obvious that Ce/La substitutions in CeRh$_6$Ge$_4$ result a rich phase diagram. Further measurements are badly required in order to establish the existence of a FM QCP induced by chemical doping and understand its unusual behavior.

	To conclude, we have studied single and polycrystalline samples of La-doped CeRh$_6$Ge$_4$. The effect of La doping is to increase the lattice volume and dilute the Kondo lattice, leading to a crossover from coherent Kondo lattice behaviors at low doping levels, to the Kondo impurity regime at higher doping. La doping also leads to suppression of the long-range magnetic order, resulting in NFL behavior. Based on the physical property measurements, we construct a phase diagram to map the evolution of the low temperature properties with La-doping. With increasing La-concentration, the ferromagnetic order is suppressed near $x_c$~=~0.26, at which pronounced strange metal behaviors with $\rho$~$\sim$~$T$ and $C/T$~$\sim$~log$(1/T)$ are observed. At higher doping (0.52~$<$~$x$~$\leq$~0.66), the NFL behavior deviates from the quantum critical behaviors of pressure-tuned CeRh$_6$Ge$_4$ \cite{13shen2020strange}, and instead can be described in terms of a disordered Kondo model. In contrast to the absence of NFL behavior in Ce$_x$La$_{1-x}$TiGe$_3$ \cite{44lee2019suppression}, the observation of strange metal behavior in La$_x$Ce$_{1-x}$Rh$_6$Ge$_4$ near the critical concentration reflects the presence of strong quantum fluctuations, indicating the need for further studies of the nature of the ground state in this region.

	\begin{acknowledgments}
		This work was supported by the National Natural Science Foundation of China (No. 12034017, and No. 11974306), the National Key R$\&$D Program of China (No. 2017YFA0303100 and No. 2016YFA0300202) and the Key R$\&$D Program of Zhejiang Province, China (2021C01002).

	\end{acknowledgments}

\end{document}